\documentclass[english]{article}
\usepackage[T1]{fontenc}
\usepackage[latin9]{inputenc}
\usepackage{geometry}
\usepackage{array}
\usepackage{booktabs}
\usepackage{amsbsy}
\usepackage{amstext}

\makeatletter

\providecommand{\tabularnewline}{\\}

\@ifundefined{date}{}{\date{}}
\makeatother

\usepackage{babel}
\begin{document}

\title{New Heuristics for Parallel and Scalable Bayesian Optimization}

\author{Ran Rubin\\
\\
Center for Theoretical Neuroscience, \\
Columbia University, New York, NY, USA}
\maketitle
\begin{abstract}
Bayesian optimization has emerged as a strong candidate tool for global
optimization of functions with expensive evaluation costs. However,
due to the dynamic nature of research in Bayesian approaches, and
the evolution of computing technology, using Bayesian optimization
in a parallel computing environment remains a challenge for the non-expert.
In this report, I review the state-of-the-art in parallel and scalable
Bayesian optimization methods. In addition, I propose practical ways
to avoid a few of the pitfalls of Bayesian optimization, such as oversampling
of edge parameters and over-exploitation of high performance parameters.
Finally, I provide relatively simple, heuristic algorithms, along
with their open source software implementations, that can be immediately
and easily deployed in any computing environment.
\end{abstract}

\section{Introduction}

Choosing a strategy for searching for the global minimum or maximum
of an unknown function is a non-trivial task. A major consideration
is the balance between the function's evaluation cost and available
resources. Different strategies are optimal for different cases. For
example, the function values may be the results of experiments that
take days or weeks to complete and only a few can be run concurrently.
In this scenario, we would like our search strategy to be as efficient
as possible, i.e maximize the function as quickly as possible by choosing
the parameters of the next function evaluation given the previous
results. In addition, we would probably be willing to invest considerable
computation resources to produce these parameters. On the other hand,
when the durations of function evaluations are only a few seconds,
new parameters must be chosen quickly which restricts the type of
computations that can be performed.

Here, we consider an intermediate case encountered in many computational
fields of science. In this case, the dimensionality of the function
is moderately high ($D\sim\mathcal{O}\left(10-100\right)$) and function
evaluations are moderately expensive ($T\sim\mathcal{O}\left(10\ \mathrm{hours}\right)$).
On the other hand, evaluating the functions usually involves some
sort of a computer simulation and current cluster computers allow
us to run many ($m\sim\mathcal{O}\left(10^{3}\right)$) function evaluations
in parallel. In addition, we assume that the optimization is done
given a particular amount of computational resources, for example
the number of CPUs we can use and the amount of time we are willing
to use them. An example that fits this scenario, from the field of
machine learning, is the training of neural networks where the parameters
to optimize are usually the various learning rates of the learning
algorithm, number of neurons and connections, etc.

What would be the requirements for a good search strategy in this
scenario? First, in high dimensions exhaustive searches are unreasonable,
thus the search must be efficient and find local extrema of the function
as quickly as possible. Second, our strategy must allow for as many
parallel evaluations of the function as our resources allow. In most
scenarios, the duration of function evaluations is variable. Thus,
it is also important that our search strategy is able to operate in
an asynchronous way, proposing new parameters as soon as each function
evaluation ends. Finally, as our computer resources continually increase
we expect to be able to evaluate $n\sim\mathcal{O}\left(10^{4}-10^{6}\right)$
parameter sets in a reasonable amount of time. Thus, our search strategy
must be scalable and able to handle such large datasets.

Several, fields of study have approached this problem and many different
proposals have been made. However, no single accepted method has been
chosen by the scientific community thus far. In addition, the software
implementations of most of the proposed algorithms are non-trivial
and many of them lack well designed, easy to use software packages.
Therefore, many practitioners opt to simply perform manual optimization,
by evaluating the function on a grid, identifying regions of interest
where performance is good and evaluating on a finer grid in these
regions. This simple approach allows for unlimited parallel evaluations,
is trivially scalable (since almost no computation is needed to choose
parameter sets to test) but is very inefficient and labor intensive. 

Many heuristic optimization algorithms, termed metaheuristics, such
as Simulated Annealing, Genetic Algorithms, Swarm Algorithms etc.,
have been proposed as viable solutions for global optimization \cite{yang_nature-inspired_2010,sun_parameter_2012}.
These approaches are usually designed for parallel evaluation of the
function and are usually scalable. However, the efficiency of their
search is unclear and comparison among them is a very challenging
task.

The field of non-smooth optimization has seen major advances with
the introduction of Generalized Pattern Search (GPS) and Mesh Adaptive
Direct Search (MADS) algorithms \cite{torczon_convergence_1997,audet_mesh_2006}.
These are fairly general frameworks for optimization that can easily
combine different search approaches, are well founded in theory and
come with convergence guaranties under mild assumptions., and can
be fairly easily parallelized and scaled. However, combining them
with efficient search approaches suitable for parallel environments
remains an active field of research \cite{acerbi_practical_2017}
with no currently available tools for the non-expert.

Finally, recent years have also seen a flourish of research in the
field of Bayesian Optimization (BO), that is, using Bayesian inference
to guide the search strategy during function optimization (for a review
see \cite{shahriari_taking_2016}). With a proper selection of prior
distributions this approach yields a very efficient search strategy.
However, Bayesian inference itself is computationally intensive which
is prohibitive in many situations. The research on how to use Bayesian
inference in a parallel environment and how to scale the inference
itself to large datasets is intense and ongoing and mature algorithms
and tools are not yet available. 

It seems that no single approach satisfies all the requirements we
set for our search strategy. In this report, I would like to examine
more closely the Bayesian approach for optimization, and review the
possible ways in which it can be parallelized and scaled. In addition,
I will present new, relatively simple, heuristic algorithms for parallel
and scalable Bayesian optimization. These algorithms combine several
recently proposed approaches and some new ideas of my own. Importantly,
I also provide an open source software implementation which is designed
to be modular and modifiable, and represents an immediately available,
working solution for moderate scale, parallel Bayesian optimization.
Hopefully, this code can serve as a foundation upon which other systematic,
large-scale approaches may be built. 

In the following, I will briefly review the state-of-the-art in Bayesian
optimization in sections 2, present a few modifications that may be
advantageous in section 3 and give the details of my proposed algorithms
in section 4.

\section{Review of Bayesian Inference and Optimization}

We are interested in finding a suitable set of parameters, $\boldsymbol{x}\in\mathcal{R}^{D}$,
that maximize some unknown function $f\left(\boldsymbol{x}\right)$.
We assume our function evaluation values, $y\left(\boldsymbol{x}\right)$,
are distributed according to,

\begin{equation}
y\left(\boldsymbol{x}\right)\sim\mathcal{N}\left(f\left(\boldsymbol{x}\right),\sigma^{2}\right)\ ,
\end{equation}

with constant, unknown standard deviation $\sigma$. 

The goal of our optimization is to find $\boldsymbol{x}^{\star}$,
such that
\begin{equation}
\boldsymbol{x}^{\star}=\arg\max_{\boldsymbol{x}}f\left(\boldsymbol{x}\right)\ .
\end{equation}

In BO we select points to evaluate $y\left(\boldsymbol{x}\right)$
by performing Bayesian inference based on past observations: We start
by adopting a probabilistic model of $f\left(\boldsymbol{x}\right)$,
which we denote as $\hat{f}\left(\boldsymbol{x}\right)$, that models
our knowledge of $f\left(\boldsymbol{x}\right)$ as a \textbf{distribution
over functions}. In particular, we choose a Gaussian Process (GP)
\cite{rasmussen_gaussian_2005} denoted as,
\begin{equation}
\hat{f}\left(\boldsymbol{x}\right)\sim\mathcal{GP}\left(\mu\left(\boldsymbol{x}\right),K\left(\boldsymbol{x},\boldsymbol{x}^{\prime}\right)\right)\ ,
\end{equation}
where $\mu\left(\boldsymbol{x}\right)$ is a mean function and $K\left(\boldsymbol{x},\boldsymbol{x}^{\prime}\right)$
is some positive definite covariance kernel. This notation implies,
that a set of points $\left\{ \boldsymbol{x}_{i}\right\} _{i=1}^{n}$
induces a multivariate normal distribution on the vector $\hat{\boldsymbol{f}}\in\mathcal{R}^{n}$
(with $\hat{f_{i}}=\hat{f}\left(\boldsymbol{x}_{i}\right)$):
\begin{equation}
\hat{\boldsymbol{f}}\sim\mathcal{N}\left(\boldsymbol{\mu},\Sigma\right)
\end{equation}
with means $\mu_{i}=\mu\left(\boldsymbol{x}_{i}\right)$ and covariance
matrix $\Sigma_{ij}=K\left(\boldsymbol{x}_{i},\boldsymbol{x}_{j}\right)$.

To perform Bayesian inference we assume some priors on the model's
parameters. First, we assume that $\hat{f}\left(\boldsymbol{x}\right)$
has a GP prior with constant mean function\footnote{Remember that the posterior distribution will not, in general, conserve
the properties of the prior distribution and therefore the constant
mean assumption is not restrictive.} $\mu\left(\boldsymbol{x}\right)=\bar{\mu}$ and a covariance kernel
$K_{\boldsymbol{\theta}}\left(\boldsymbol{x},\boldsymbol{x}^{\prime}\right),$
with a set of hyper-parameters $\boldsymbol{\theta}$: 
\begin{equation}
\hat{f}\left(\boldsymbol{x}\right)\sim\mathcal{GP}\left(\bar{\mu},K_{\boldsymbol{\theta}}\left(\boldsymbol{x},\boldsymbol{x}^{\prime}\right)\right)\ .
\end{equation}
Further, we specify priors for the the noise magnitude, GP prior mean
and the kernel's hyper-parameters: 
\begin{equation}
\sigma\sim P\left(\sigma\right),\,\bar{\mu}\sim P\left(\bar{\mu}\right),\,\boldsymbol{\theta}\sim P\left(\boldsymbol{\theta}\right)\ .
\end{equation}

The choice of kernel $K_{\boldsymbol{\theta}}\left(\boldsymbol{x},\boldsymbol{x}^{\prime}\right)$
is an important step, as different kernels confer different properties
on the inferred functions and may be suitable in different situations
\cite{rasmussen_gaussian_2005}. The kernel hyper-parameters, $\boldsymbol{\theta}$,
usually represent, the typical amplitude and length scales of our
inferred functions. 

Now, given a set of $n$ observation pairs, 
\begin{equation}
O=\left\{ \boldsymbol{x}_{i},y_{i}\right\} _{i=1}^{n}\ ,
\end{equation}
we use Bayes rule to infer the posterior distribution of the parameters:
\begin{equation}
P\left(\hat{f}\left(\boldsymbol{x}\right),\sigma,\bar{\mu},\boldsymbol{\theta}|O\right)\propto P\left(O|\hat{f}\left(\boldsymbol{x}\right),\sigma\right)P\left(\sigma\right)P\left(\hat{f}\left(\boldsymbol{x}\right)|\bar{\mu},\boldsymbol{\theta}\right)P\left(\bar{\mu}\right)P\left(\boldsymbol{\theta}\right)
\end{equation}

The full posterior distribution lets us make predictions on $f\left(\boldsymbol{x}\right)$
and also provides us with an estimate of the level of certainty of
our predictions. One attractive feature of a GP is that, given the
parameters $\sigma,\bar{\mu}$ and $\boldsymbol{\theta}$ the posterior
distribution of functions, $\hat{f}\left(\boldsymbol{x}\right)$,
given $O$ is also a GP and it is possible to calculate analytically
its posterior mean function and covariance kernel, simplifying calculations
involving the full posterior distribution (for details see \cite{rasmussen_gaussian_2005}).

\subsection{Scalability of Bayesian Inference}

Due to the required inversion of the observations' covariance matrix,
full Bayesian inference for GPs scales as $\mathcal{O}(n^{3})$ with
the number of observations, $n$. To be useful in higher dimensions,
when $10^{3}-10^{6}$ of observations are needed to explore the parameter
space, we must use an approximation method that scales as $\mathcal{O}(n)$.
The leading solution in the literature is variational inference (VI)
\cite{blei_variational_2017}, which for GPs amounts to using the
kernel method to approximate the posterior mean and covariance functions
with a fixed number of kernel elements. A recent proposal \cite{cheng_variational_2017}
proposes a VI algorithm that indeed scales as $\mathcal{O}(n)$. The
improved scaling comes at the cost of a somewhat reduced expressibility
and an overestimation of the predicted variance. An alternative approach
for the scaling of Bayesian inference is to reduce the number of observations
used in the inference by considering only local Bayesian inference
around promising loci in the parameter space \cite{acerbi_practical_2017}
or by performing random sub-sampling of the observations \cite{pakman_stochastic_2016}.

\subsection{Bayesian Search Strategies}

Given a set of observation, $O$, a BO search strategy proposes a
new point $\boldsymbol{x}_{\text{new}}$ according to some criteria.
This is usually done by choosing the point that maximizes some function
over the search domain, termed the acquisition function. 

Although many acquisition functions have been proposed (see \cite{shahriari_taking_2016}
and references therein), a common one is the expected improvement
(EI) which we define here as:
\begin{equation}
EI\left(\boldsymbol{x}\right)=E_{P\left(\hat{f}\left(\boldsymbol{x}\right),\sigma,\bar{\mu},\boldsymbol{\theta}|O\right)}\left(\left[\hat{f}\left(\boldsymbol{x}\right)-\max_{i}\hat{f}\left(\boldsymbol{x}_{i}\right)\right]_{+}\right)\ ,
\end{equation}
where the expectation is taken over the posterior distribution given
the observations, and $\left[z\right]_{+}=z$ for $z>0$ and $\left[z\right]_{+}=0$
otherwise. We then choose:
\begin{equation}
\boldsymbol{x}_{\text{new}}^{EI}=\arg\max_{\boldsymbol{x}}EI\left(\boldsymbol{x}\right)\ .
\end{equation}
We note that the EI can be high either because the actual value of
the posterior mean, $\mu_{O}\left(\boldsymbol{x}\right)$, is high
or the uncertainty (posterior variance, $\sigma_{O}^{2}\left(\boldsymbol{x}\right)$)
at point $\boldsymbol{x}$ is high. Thus, this measure implements
a kind of exploration-exploitation balance. 

For a GP the EI can be relatively easily calculated following a few
simple approximations: First, practitioners remove the sample dependence
of the max operation by replacing $\hat{f}\left(\boldsymbol{x}_{i}\right)$
with the posterior mean, $\mu_{O}\left(\boldsymbol{x}_{i}\right)$,
or with the observations themselves, $y_{i}$. Then, the EI can be
calculated analytically conditioned on $\sigma,$ $\bar{\mu}$ and
$\boldsymbol{\theta}$ and practitioners usually approximate the expectancy
over these parameters by using a periodically updated point estimate
for them or by averaging a set of samples from their posterior using
Markov Chain Monte Carlo (MCMC) methods (which usually significantly
increase the computational cost of the inference). 

Finally, we note that maximizing the EI (or any other acquisition
function) is a non-trivial task in itself, and may be computationally
expensive and require some form of approximation.

\subsection{Parallel Bayesian Search Strategies}

Since Bayesian inference is computationally intensive, even a relatively
short inference time (for example, a few seconds) may create a computational
bottleneck if a large number of inferences is required. This may leave
some of our computer resources idle while waiting for the new points
to evaluate. As proposed in \cite{hernandez-lobato_parallel_2017},
one way to make sure that the inference remains scalable is to perform
the Bayesian inference itself in parallel. However, a couple of difficulties
arise in this case:

First, computing several points for evaluation in parallel, given
the same or very similar sets of known measurements, will typically
result in proposing the same, or very similar, new point, which will
hinder the exploration of the function's parameters space. One way
to avoid this problem is to introduce variability in the proposed
new points. Recently, the use of Thompson Sampling \cite{thompson_likelihood_1933,thompson_theory_1935,russo_tutorial_2017}
has been proposed \cite{hernandez-lobato_parallel_2017,kandasamy_parallelised_2018}
as a tool for generating this variability in a way that is consistent
with the posterior distribution. Given observations, $O$, and a single
sample of $\hat{f}\left(\boldsymbol{x}\right),\sigma,\bar{\mu},$
and $\boldsymbol{\theta}$ from the posterior distribution, Thompson
sampling proposes a new point for evaluations according to
\begin{equation}
\boldsymbol{x}_{\text{new}}^{\mathrm{Thompson}}=\arg\max_{\boldsymbol{x}}\hat{f}\left(\boldsymbol{x}\right)\ .
\end{equation}
Simply put, Thompson Sampling can be regarded as a probability matching
search strategy, setting the probability of choosing a point $\boldsymbol{x}$
to the posterior probability this point is indeed the maximum of $f\left(\boldsymbol{x}\right)$.
When several inferences run in parallel, each will sample from the
posterior probability of $\boldsymbol{x}^{\star}$ independently,
and may therefore propose different points for evaluation. We note
that, although conceptually simple, sampling $\hat{f}\left(\boldsymbol{x}\right)$
from the posterior and maximizing over it is an additional computational
step that may not be trivial to implement in a scalable way

Second, performing inference in an asynchronous parallel environment
implies that in addition to the set of observations, there is an additional,
possibly significant, set of running experiments/simulations for which
the parameter values are known but their results are still pending.
Snoek et al. \cite{snoek_practical_2012} propose to take this second
set into account during the inference by 'fantasizing' the results
of the evaluation and performing the inference with the fantasized
values. 

It is simple to incorporate fantasies within the framework of Thompson
Sampling. Given observations, $O$, and pending points $P=\left\{ \boldsymbol{x}_{i}^{\mathrm{Pending}}\right\} $,
one first samples fantasized observation values, $\tilde{y_{i}}\sim\mathcal{N}\left[\hat{f}\left(\boldsymbol{x}_{i}^{\mathrm{Pending}}\right),\sigma\right]$
where $\hat{f}$ and $\sigma$ are sampled from the posterior distribution
given the observations. Then, another $\hat{f}\left(\boldsymbol{x}\right)$
can be sampled from the posterior distribution given the union $O\cup\left\{ \boldsymbol{x}_{i}^{\mathrm{pending}},\tilde{y}_{i}\right\} $.
While this procedure will not, on average, change the predicted mean
of the function at the pending points, it will reduce the uncertainty
around points that are pending and make them less likely to be selected
again. In addition, as noted in \cite{hernandez-lobato_parallel_2017},
sampling $\tilde{y_{i}}$ from the posterior distribution does not,
on average, change the posterior of the GP hyper-parameters, so there
is no need to sample them again conditioned on $\tilde{y_{i}}$.

\section{Practical Considerations and Heuristic Modifications for Improved
Efficiency}

Until now I have reviewed the state-of-the-art of parallel and scalable
Bayesian optimization. Here I propose three modifications and improvements
that may increase the efficiency of the Bayesian optimization. I found
these to be helpful in my experiments but did not conduct a systematical
quantification of their effect. Such a quantification is not a trivial
task and is beyond the scope of this technical report. Nonetheless,
I present them here as possible, fruitful avenues of research. 

\subsection{Sample Improvement }

Thompson Sampling has been shown to bound the mean regret when the
goal of optimization is to maximize the accumulative value or minimize
the accumulated regret over time \cite{agrawal_thompson_2013,kaufmann_thompson_2012,agrawal_further_2013}.
As such, it tends to lean more on exploitation once a `good` set of
parameters is found. Indeed, it has been claimed that Thompson sampling
asymptotically converges to the global maximizer in exponential time
\cite{basu_analysis_2017}. However, in the context of parameter optimization,
there is really no need to evaluate the same point again once we are
confident enough about the value of $f\left(\boldsymbol{x}\right)$
at this point.

Here I propose the concept of Sample Improvement (SI). We start with
the simple idea to approximate/replace the expected improvement by
a single sample of the improvement, and select the point $\boldsymbol{x}$
that maximizes this sample improvement function. Given observations,
$O$, and a single sample of $\hat{f}\left(\boldsymbol{x}\right),\sigma,\bar{\mu},$
and $\boldsymbol{\theta}$ from the posterior distribution, we define
the Sample Improvement of point $\boldsymbol{x}$, as simply
\begin{equation}
SI\left(\boldsymbol{x}\right)=\left[\hat{f}\left(\boldsymbol{x}\right)-\max_{i}\hat{f}\left(\boldsymbol{x}_{i}\right)\right]_{+}\ .
\end{equation}
As an approximation, the SI will introduce the variability that we
can exploit for parallel inference. We therefore choose the new point
for evaluation according to
\begin{equation}
\boldsymbol{x}_{\text{new}}^{SI}=\arg\max_{\boldsymbol{x}}SI\left(\boldsymbol{x}\right)\ .
\end{equation}

Note that the expected improvement is always positive so the $\arg\max$
operation is a well-defined function. However, in case no point in
the sample improves the result of the function we are left with $SI\left(\boldsymbol{x}\right)=0$
for all $\boldsymbol{x}$ and the $\arg\max$ is not defined. The
advantage of using SI over Thompson sampling (Eq. 11) is that we can
interpret a sample function for which $SI\left(\boldsymbol{x}\right)\le\epsilon$
for all $\boldsymbol{x}$ and some a priori $\epsilon>0$ as a sign
of possible convergence of the inference to a local minimum. In this
case, it might be appropriate to use a more exploratory strategy for
the selection of points for evaluation. One possibility is choosing
the point that locally maximizes the estimated variance or an upper
confidence bound of $f\left(\boldsymbol{x}\right)$ (This will not
generate variability in the selected point which may be problematic
if many inferences are run in parallel). Another is choosing random
points around the current estimated maximum to encourage exploration
in its vicinity.

\subsection{Variance Control}

Alternating between search strategies to improve the exploration-exploitation
balance based on the current state of the search is important for
parameter optimization. Indeed several meta-strategies were proposed
to select between the sampling proposals of several strategies \cite{brochu_portfolio_2010,shahriari_entropy_2014}.
However, finding a strategy that will be suitable at every stage of
the optimization and that will always prevent oversampling of similar
points remains a challenge. Here I propose a simple strategy, termed
Variance Control, to ensure that no point is ever oversampled by explicitly
controlling the estimated variance of sampled points. We consider
a situation in which we wish to evaluate $f\left(\boldsymbol{x}\right)$
at any given point, only up to some level of accuracy. Bayesian inference
can aid in setting this level: A reasonable limit might be a fraction
of the estimated noise level in our observation ($\sigma^{2}$). We
would like our algorithm to avoid sampling points where the estimated
variance is below our required accuracy. This can be achieved by excluding
these points from our search domain. 

Given a set of observations $O=\left\{ \boldsymbol{x}_{i},y_{i}\right\} $
and pending points $P=\left\{ \boldsymbol{x}_{i}^{\mathrm{Pending}}\right\} $
and a sample of $\sigma,\bar{\mu},$ and $\boldsymbol{\theta}$, we
can define our search domain
\begin{equation}
D_{O,P}=\left\{ \boldsymbol{x}|\sigma_{O,P}\left(\boldsymbol{x}\right)>\rho\sigma\right\} \ ,
\end{equation}
where $\sigma_{O,P}^{2}\left(\boldsymbol{x}\right)$ is the posterior
estimated variance given $\sigma,\bar{\mu},$ and $\boldsymbol{\theta}$,
and $\rho$ is our desired level of accuracy as a fraction of the
estimated noise. In this setting, we can view parameter optimization
not as an attempt to find the maximum of $f\left(\boldsymbol{x}\right)$
over the entire parameter space but rather only over $D_{O,P}$. Therefore,
we choose our next point to evaluate according to our current search
strategy only if the proposed point is in $D_{O,P}$. In addition,
we also compute improvement measures based only on observations that
are in $D_{O,P}$. We will describe two different approaches for implementing
variance control, in section 4.3.

\subsection{Boundary Avoidance}

In many texts, examples of Bayesian inference are presented for a
function of one or two parameters. While this is instructive, one
has to remember that our intuitive understanding of low-dimensional
spaces does not always work well for higher dimensions. A famous example
is the volume of a $D$ dimensional sphere: unlike 2-dim. circles
or 3-dim. balls, when $D\gg1$ most of the volume of a sphere is concentrated
at the edge of the sphere. 

In the context of Bayesian optimization, we are usually concerned
with finding the best parameters within some range for each parameter.
The range is usually picked based on our prior belief of the reasonable
value of the parameters. In fact, we usually suspect that the optimal
value is not close to the edges of our range. However, when choosing
points to sample using Bayesian inference, we are very likely to choose
points on the edges because, unless already sampled, at these point
the model is extrapolating and we expect to have large expected variance
and, possibly, a monotonically increasing mean function. For functions
of one or two parameters, evaluating the function at the edges of
our range will only take a few samples and will quickly reduce the
estimated variance there. However for higher dimensional spaces, sampling
the edges well will require many function evaluations in locations
that we do not believe are optimal. 

Taking the Bayesian approach, we would like to express our knowledge
of the parameter range as a prior distribution on the location of
the maximum of $f\left(\boldsymbol{x}\right)$. However, since we
only directly model our knowledge of $f\left(\boldsymbol{x}\right)$
and not of the location of its maximum, it is not clear how to incorporate
this prior within the GP framework we described. First steps at systematically
addressing the edges oversampling problem were taken in \cite{siivola_correcting_2017}.
Our practical solution, is to simply reject points for evaluations
if they are on the edges of our parameter range.

\section{Methods and Algorithms}

Implementing Bayesian inference and optimization involves many choices
and approximations that may affect the results. However, it is very
hard to estimate systematically the effect of these choices, and the
implementer must rely on choices previously made by others as well
as common sense and intuition. In this section, I would like to inform
the reader and users of these choices and approximations so that proper
review and improvement can be made. In addition, I have tried to design
my algorithmic implementation to be as modular as possible, so that
various parts can be easily replaced in case some choice or approximation
is not to the user's liking or is unsuitable for the user's circumstances. 

As an immediately available working solution, I provide two simple
algorithms that perform parallel, Bayesian inference using the above
proposed heuristics to generate variability in the selection of points,
avoid boundaries, alternate between global and local search strategies
and avoid oversampling of points. An open source software implementation
of these algorithms along with further implementation details, can
be found in the GitHub repository: https://github.com/ranr01/miniBOP.

\subsection{Parallel Bayesian Optimization}

This algorithm assumes that each computational node can be used to
either evaluate the function at a single set of parameters, $\boldsymbol{x}$,
or choose a new point for evaluation using Bayesian inference. Given
$m$ computational nodes, a simple algorithm to perform the optimization
can be defined as follows:
\begin{enumerate}
\item Submit a batch of $m$ function evaluations on a quasi-random grid. 
\item Wait for a job to finish.
\begin{enumerate}
\item If the finished job is function evaluation, process results and submit
a Bayesian inference job with the current information of pending and
completed simulations.
\item If the finished job is a Bayesian inference job, submit a function
evaluation with the suggested parameters. 
\end{enumerate}
\item Repeat 2 until maximal amount of resources are used. 
\end{enumerate}
For a more detailed discussion of different resource management strategies
I refer the reader to \cite{jones_taxonomy_2001,hutter_parallel_2012}.

\subsection{Sampling and maximizing a function from the posterior GP}

Sampling a function $\hat{f}\left(\boldsymbol{x}\right)$ from the
posterior GP is done by sampling a multivariate Gaussian with the
appropriate mean and covariance according to the sampled points. This
can be done in batch by computing the Cholesky decomposition of the
full covariance matrix of the sampled points. For sequential sampling
of points (as is usually required for maximization) the Cholesky decomposition
can be performed iteratively in an efficient manner. 

Note that, this procedure does not scale well with the number of sampled
points from $\hat{f}\left(\boldsymbol{x}\right)$. Finding an efficient,
scalable way of sampling from a high-dimensional multivariate Gaussian
is in fact an open question under scientific investigation (see for
example \cite{simpson_scalable_2013,ambikasaran_fast_2014}). To avoid
sampling a large number of points from a single sampled function,
during our inference we do not try to find a global maximum of $\hat{f}\left(\boldsymbol{x}\right)$.
Instead, we find a local maximum by running the Nelder-Mead (Simplex)
algorithm starting from a random initial point. We specify an absolute
tolerance for $\boldsymbol{x}$ as a tunable parameter.

\subsection{Choosers}

The heart of Bayesian optimization is the algorithm to choose new
points for evaluation. Here we present two heuristic algorithms that
incorporate the principles discussed above.

\subsubsection{Bayesian Optimization and Poll steps (BOP)}

This algorithm alternates between approximately maximizing the Sample
Improvement and taking random steps from the current global maximum.
The name 'poll steps' is borrowed from the GPR and MADS algorithm
to describe the random exploration around the current estimated maximum.

Given a set of observations $O=\left\{ \boldsymbol{x}_{i},y_{i}\right\} $
and pending points $P=\left\{ \boldsymbol{x}_{i}^{\mathrm{Pending}}\right\} $
we choose the next point in the following way:
\begin{enumerate}
\item Sample the hyper-parameters ($\sigma,\bar{\mu},\ \mathrm{and}\ \boldsymbol{\theta}$)
from the posterior given the observation set $O$, using MCMC.
\item Sample the results of the pending points by sampling $\tilde{y_{i}}\sim\mathcal{N}\left[\hat{f}\left(\boldsymbol{x}_{i}^{\mathrm{Pending}}\right),\sigma\right]$
were $\hat{f}\left(\boldsymbol{x}\right)$ is sampled from the posterior
GP given the observations set $O$.
\item Sample a set of candidate points, $C_{\mathrm{Bayes}}=\left\{ \boldsymbol{x}_{j}^{\mathrm{cand}},\hat{f}_{j}\right\} _{j=1}^{n_{\mathrm{cand}}}$,
$\hat{f}_{j}=\hat{f}\left(\boldsymbol{x}_{j}^{\mathrm{cand}}\right)$,
by finding a local maximum of a sample function $\hat{f}\left(\boldsymbol{x}\right)$,
sampled from the posterior GP given the union $O\cup\left\{ \boldsymbol{x}_{i}^{\mathrm{pending}},\tilde{y}_{i}\right\} $.
To avoid sampling a large number of points from a single sample function
we sample a new $\hat{f}\left(\boldsymbol{x}\right)$ for each $\boldsymbol{x}_{j}^{\mathrm{cand}}$.
\item Exclude candidate points in $C_{\mathrm{Bayes}}$ that have low predicted
variance (i.e. $\boldsymbol{x}_{j}^{\mathrm{cand}}\notin D_{O,P}$)
or that are on the edges of the parameters range.
\item Calculate the improvement of the remaining points, where improvement
is defined as
\begin{equation}
I_{j}=\left[\hat{f}_{j}-\max_{\boldsymbol{x}_{i}\in O,P}\mu_{O,P}\left(\boldsymbol{x}_{i}\right)\right]_{+}
\end{equation}
\item Exclude candidate points in $C_{\mathrm{Bayes}}$ for which the improvement
is zero (or smaller than some value $\epsilon$).
\item If $C_{\mathrm{Bayes}}$ is not empty return the point with the maximal
improvement and exit.
\item Poll step: 
\begin{enumerate}
\item Find the best parameters out of the sampled and pending points.
\item Choose a set of candidate points, $C_{\mathrm{poll}}=\left\{ \boldsymbol{x}_{j}^{\mathrm{poll}}\right\} _{j=1}^{n_{\mathrm{poll}}}$
by choosing a random point around the best point. The random step
size is chosen to be proportional to the estimated length scales of
the covariance kernel.
\item Exclude points with low predicted variance.
\item if $C_{\mathrm{poll}}$ is not empty, return the point with the maximal
predicted variance and exit.
\end{enumerate}
\item Default step: return a random point and exit.
\end{enumerate}

\subsubsection{Function Barrier Variance Control (FuBar VC)}

Here, we approximate the requirement of sampling points only inside
$D_{O,P}$ by adding a soft barrier function to our sampled function
$\hat{f}\left(\boldsymbol{x}\right)$. We define
\begin{equation}
g\left(\boldsymbol{x}\right)=\hat{f}\left(\boldsymbol{x}\right)-b\left(\sigma_{O,P}\left(\boldsymbol{x}\right)\right)
\end{equation}
where $\sigma_{O,P}\left(\boldsymbol{x}\right)$ is the estimated
standard deviation of $f\left(\boldsymbol{x}\right)$ at the point
\textbf{$\boldsymbol{x}$} and $b\left(s\right)$ is a function which
is very large for $s<\rho\sigma$ and negligible for $s>\rho\sigma$.
As a concrete example, we choose a power law function, 
\begin{equation}
b\left(s\right)=\left(\frac{\rho\sigma}{s}\right)^{z}\ ,
\end{equation}
with $z=10$. Here, $z$ controls how sharp our barrier function is
around the boundary $\sigma_{O,P}\left(\boldsymbol{x}\right)=\rho\sigma$.
This parameter can control the rate of exploration around the boundary
as higher values allow sampling of points closer to the boundary.

The resulting algorithm remains the same as the BOP algorithm except
the two following modifications:
\begin{itemize}
\item In step 3 we replace the optimization of $\hat{f}\left(\boldsymbol{x}\right)$
with the optimization of $g\left(\boldsymbol{x}\right)$ 
\item In step 5 improvement is defined as 
\begin{equation}
I_{j}=\left[g_{j}-\max_{\boldsymbol{x}_{i}\in O,P}\left[\mu_{O,P}\left(\boldsymbol{x}_{i}\right)-b\left(\sigma_{O,P}\left(\boldsymbol{x}_{i}\right)\right)\right]\right]_{+}
\end{equation}
\end{itemize}
We note that using the barrier function renders the estimated variance
check in step 4 somewhat redundant, so in practice it can be dropped.

\subsection{Other implementation details}

\subsubsection{Parameter rescaling}

We consider a scenario in which optimization is performed in a hyper-box
specified by a minimal and maximal value for each function variable.
For the purpose of Bayesian inference, we follow the common practice
of rescaling all the variables to lie inside the unit hyper-cube,
i.e. inside the $\left[0,1\right]$ range.

\subsubsection{Covariance kernel function}

Following \cite{snoek_practical_2012} we use the Matern 5/2 kernel
with individual length scale for each dimension (sometimes termed
Automatic Relevance Determination).

Thus the kernel has $D+1$ hyper-parameters, $\left\{ \theta_{k}\right\} _{k=0}^{D}$,
and given $\boldsymbol{x}_{i}$ and $\boldsymbol{x}_{j}$ we have:
\begin{equation}
K_{\boldsymbol{\theta}}\left(\boldsymbol{x}_{i},\boldsymbol{x}_{j}\right)=\theta_{0}^{2}\left(1+\sqrt{5}r+\frac{5}{3}r^{2}\right)\exp\left(-\sqrt{5}r\right)\ ,
\end{equation}
with, 
\begin{equation}
r=\sqrt{\sum_{k=1}^{D}\left(\frac{x_{i}^{k}-x_{j}^{k}}{\theta_{k}}\right)^{2}}\ .
\end{equation}

\subsubsection{MCMC for hyper-parameter sampling}

We sample the hyper-parameters ($\sigma,\bar{\mu},\ \mathrm{and}\ \boldsymbol{\theta}$)
from the posterior given the observations by performing $T_{\mathrm{MCMC}}$
MCMC steps using Slice Sampling in the manner done in \cite{snoek_practical_2012}.

\subsubsection{Assumed Priors}

We specify prior distributions for $\bar{\mu},$ $\sigma^{2}$ and
$\boldsymbol{\theta}$.

For the mean of the GP we take a flat prior
\begin{equation}
\bar{\mu}\sim U\left(\bar{\mu}_{\min},\bar{\mu}_{\max}\right)
\end{equation}
where $U\left(a,b\right)$ is a uniform distribution between \textbf{$a$}
and $b$. We set $\bar{\mu}_{\min(\max)}$ to be the min(max) over
all function value observations.

Following \cite{snoek_practical_2012}, for the observation noise
variance we take a probability distribution 
\begin{equation}
P_{\mathrm{noise}}\left(\sigma^{2}\right)\propto\log\left[1+\left(\frac{v_{\mathrm{noise}}}{\sigma^{2}}\right)^{2}\right]
\end{equation}
and, for the GP amplitude, $\theta_{0}^{2}$ we take a log-normal
distribution 
\begin{equation}
P_{\mathrm{amp}}\left(\theta_{0}^{2}\right)\propto\frac{1}{\theta_{0}^{2}}\exp\left[-\frac{1}{2}\left(\frac{\log\left(\theta_{0}^{2}\right)}{A^{2}}\right)^{2}\right]
\end{equation}
For each length scale parameter, we use an inverse Gamma distribution:
\begin{equation}
P_{\mathrm{length}}\left(\theta_{k}\right)\propto\theta_{k}^{-\left(\alpha_{\mathrm{length}}+1\right)}\exp\left(-\frac{\lambda_{\mathrm{length}}}{\theta_{k}}\right)\ .
\end{equation}

See source code for the chosen default values of $v_{\mathrm{noise}}$,
$A^{2}$, $\alpha_{\mathrm{length}}$ and $\lambda_{\mathrm{length}}$. 

\section{Discussion}

We set out to find a search strategy suitable for global optimization
of expensive functions in a massively-parallel computing environment.
As such we needed a strategy that would be efficient, parallel and
scalable. Unfortunately, most of the available open source software
for optimization was not explicitly designed for this parallel scenario
(see \cite{shahriari_taking_2016} for a list of some of the available
software). One proprietary software solution for massive Bayesian
optimization \cite{golovin_google_2017} exists. However, its opaque
methods and algorithms make it hard to evaluate its performance. 

In my opinion, the combination of quick Bayesian search methods (such
as variational inference) with parallel versions of mathematically
grounded grid search methods such as MADS is a probable way forward
to produce robust and scalable methods of Bayesian Optimization for
large-scale global optimization. First steps in this direction have
been taken with the recently proposed BADS algorithm \cite{acerbi_practical_2017}. 

The heuristic algorithms I presented here perform efficient Bayesian
search in a parallel environment. The scalability of these algorithms
is limited only by the scalability of the Bayesian inference and sampling.
In the current software implementation, the classic full Bayesian
inference method is used, scaling $\mathcal{O}\left(n^{3}\right)$
with the number of observed points. This practically limits the application
of the algorithms moderate size problems where to only a few thousands
of observed point are needed. Once a scalable implementation of Bayesian
inference and sampling is available, both algorithms presented here
can be scaled as well. As mentioned above, a systematic quantification
of the performance of these algorithms and a thorough comparison to
other approaches is still lacking. Hopefully, they can serve as a
useful tool until such a study can be performed.

\subsection*{Acknowledgments}

I thank Larry Abbott, James Murray, Fabio Stefanini and Ari Pakman
for helpful discussions and comments. This work was supported by the
Center for Theoretical Neuroscience's NeuroNex award and by the charitable
contribution of the Gatsby Foundation. 

\bibliographystyle{unsrt}
\bibliography{BayesianOptimization}

\begin{thebibliography}{10}

\bibitem{yang_nature-inspired_2010}
Xin-She Yang.
\newblock {\em Nature-{{Inspired Metaheuristic Algorithms}}: {{Second
  Edition}}}.
\newblock {Luniver Press}, 2010.

\bibitem{sun_parameter_2012}
Jianyong Sun, Jonathan~M. Garibaldi, and Charlie Hodgman.
\newblock Parameter {{Estimation Using Metaheuristics}} in {{Systems Biology}}:
  {{A Comprehensive Review}}.
\newblock {\em IEEE/ACM Trans. Comput. Biol. Bioinformatics}, 9(1):185--202,
  January 2012.

\bibitem{torczon_convergence_1997}
V.~Torczon.
\newblock On the {{Convergence}} of {{Pattern Search Algorithms}}.
\newblock {\em SIAM Journal on Optimization}, 7(1):1--25, February 1997.

\bibitem{audet_mesh_2006}
Charles Audet and John~E. Dennis~Jr.
\newblock Mesh adaptive direct search algorithms for constrained optimization.
\newblock {\em SIAM Journal on optimization}, 17(1):188--217, 2006.

\bibitem{acerbi_practical_2017}
Luigi Acerbi and Wei Ji.
\newblock Practical {{Bayesian Optimization}} for {{Model Fitting}} with
  {{Bayesian Adaptive Direct Search}}.
\newblock In I.~Guyon, U.~V. Luxburg, S.~Bengio, H.~Wallach, R.~Fergus,
  S.~Vishwanathan, and R.~Garnett, editors, {\em Advances in {{Neural
  Information Processing Systems}} 30}, pages 1836--1846. {Curran Associates,
  Inc.}, 2017.

\bibitem{shahriari_taking_2016}
B.~Shahriari, K.~Swersky, Z.~Wang, R.~P. Adams, and N.~de~Freitas.
\newblock Taking the {{Human Out}} of the {{Loop}}: {{A Review}} of {{Bayesian
  Optimization}}.
\newblock {\em Proceedings of the IEEE}, 104(1):148--175, January 2016.

\bibitem{rasmussen_gaussian_2005}
Carl~Edward Rasmussen and Christopher K.~I. Williams.
\newblock {\em Gaussian {{Processes}} for {{Machine Learning}}}.
\newblock {The MIT Press}, Cambridge, Mass, November 2005.

\bibitem{blei_variational_2017}
David~M. Blei, Alp Kucukelbir, and Jon~D. McAuliffe.
\newblock Variational {{Inference}}: {{A Review}} for {{Statisticians}}.
\newblock {\em Journal of the American Statistical Association},
  112(518):859--877, April 2017.

\bibitem{cheng_variational_2017}
Ching-An Cheng and Byron Boots.
\newblock Variational {{Inference}} for {{Gaussian Process Models}} with
  {{Linear Complexity}}.
\newblock {\em arXiv:1711.10127 [cs, stat]}, November 2017.

\bibitem{pakman_stochastic_2016}
Ari Pakman, Dar Gilboa, David Carlson, and Liam Paninski.
\newblock Stochastic {{Bouncy Particle Sampler}}.
\newblock {\em arXiv:1609.00770 [stat]}, September 2016.

\bibitem{hernandez-lobato_parallel_2017}
Jos{\'e}~Miguel Hern{\'a}ndez-Lobato, James Requeima, Edward~O. Pyzer-Knapp,
  and Al{\'a}n Aspuru-Guzik.
\newblock Parallel and {{Distributed Thompson Sampling}} for {{Large}}-scale
  {{Accelerated Exploration}} of {{Chemical Space}}.
\newblock {\em arXiv:1706.01825 [stat]}, June 2017.

\bibitem{thompson_likelihood_1933}
William~R. Thompson.
\newblock {{ON THE LIKELIHOOD THAT ONE UNKNOWN PROBABILITY EXCEEDS ANOTHER IN
  VIEW OF THE EVIDENCE OF TWO SAMPLES}}.
\newblock {\em Biometrika}, 25(3-4):285--294, December 1933.

\bibitem{thompson_theory_1935}
William~R. Thompson.
\newblock On the {{Theory}} of {{Apportionment}}.
\newblock {\em American Journal of Mathematics}, 57(2):450--456, 1935.

\bibitem{russo_tutorial_2017}
Daniel Russo, Benjamin Van~Roy, Abbas Kazerouni, Ian Osband, and Zheng Wen.
\newblock A {{Tutorial}} on {{Thompson Sampling}}.
\newblock {\em arXiv:1707.02038 [cs]}, July 2017.

\bibitem{kandasamy_parallelised_2018}
Kirthevasan Kandasamy, Akshay Krishnamurthy, Jeff Schneider, and Barnabas
  Poczos.
\newblock Parallelised {{Bayesian Optimisation}} via {{Thompson Sampling}}.
\newblock In {\em International {{Conference}} on {{Artificial Intelligence}}
  and {{Statistics}}}, pages 133--142, March 2018.

\bibitem{snoek_practical_2012}
Jasper Snoek, Hugo Larochelle, and Ryan~P Adams.
\newblock Practical {{Bayesian Optimization}} of {{Machine Learning
  Algorithms}}.
\newblock In F.~Pereira, C.~J.~C. Burges, L.~Bottou, and K.~Q. Weinberger,
  editors, {\em Advances in {{Neural Information Processing Systems}} 25},
  pages 2951--2959. {Curran Associates, Inc.}, 2012.

\bibitem{agrawal_thompson_2013}
Shipra Agrawal and Navin Goyal.
\newblock Thompson sampling for contextual bandits with linear payoffs.
\newblock In {\em International {{Conference}} on {{Machine Learning}}}, pages
  127--135, 2013.

\bibitem{kaufmann_thompson_2012}
Emilie Kaufmann, Nathaniel Korda, and R{\'e}mi Munos.
\newblock Thompson sampling: {{An}} asymptotically optimal finite-time
  analysis.
\newblock In {\em International {{Conference}} on {{Algorithmic Learning
  Theory}}}, pages 199--213. {Springer}, 2012.

\bibitem{agrawal_further_2013}
Shipra Agrawal and Navin Goyal.
\newblock Further optimal regret bounds for thompson sampling.
\newblock In {\em Artificial {{Intelligence}} and {{Statistics}}}, pages
  99--107, 2013.

\bibitem{basu_analysis_2017}
Kinjal Basu and Souvik Ghosh.
\newblock Analysis of {{Thompson Sampling}} for {{Gaussian Process
  Optimization}} in the {{Bandit Setting}}.
\newblock {\em arXiv:1705.06808 [stat]}, May 2017.

\bibitem{brochu_portfolio_2010}
Eric Brochu, Matthew~W. Hoffman, and Nando {de Freitas}.
\newblock Portfolio {{Allocation}} for {{Bayesian Optimization}}.
\newblock {\em arXiv:1009.5419 [cs]}, September 2010.

\bibitem{shahriari_entropy_2014}
Bobak Shahriari, Ziyu Wang, Matthew~W. Hoffman, Alexandre
  Bouchard-C{\^o}t{\'e}, and Nando {de Freitas}.
\newblock An {{Entropy Search Portfolio}} for {{Bayesian Optimization}}.
\newblock {\em arXiv:1406.4625 [cs, stat]}, June 2014.

\bibitem{siivola_correcting_2017}
Eero Siivola, Aki Vehtari, Jarno Vanhatalo, and Javier Gonz{\'a}lez.
\newblock Correcting boundary over-exploration deficiencies in {{Bayesian}}
  optimization with virtual derivative sign observations.
\newblock {\em arXiv:1704.00963 [stat]}, April 2017.

\bibitem{jones_taxonomy_2001}
Donald~R. Jones.
\newblock A {{Taxonomy}} of {{Global Optimization Methods Based}} on {{Response
  Surfaces}}.
\newblock {\em Journal of Global Optimization}, 21(4):345--383, December 2001.

\bibitem{hutter_parallel_2012}
Frank Hutter, Holger~H. Hoos, and Kevin Leyton-Brown.
\newblock Parallel {{Algorithm Configuration}}.
\newblock In {\em Learning and {{Intelligent Optimization}}}, Lecture Notes in
  Computer Science, pages 55--70. {Springer, Berlin, Heidelberg}, 2012.

\bibitem{simpson_scalable_2013}
Daniel~P. Simpson, Ian~W. Turner, Christopher~M. Strickland, and Anthony~N.
  Pettitt.
\newblock Scalable iterative methods for sampling from massive {{Gaussian}}
  random vectors.
\newblock {\em arXiv:1312.1476 [math, stat]}, December 2013.

\bibitem{ambikasaran_fast_2014}
Sivaram Ambikasaran, Daniel Foreman-Mackey, Leslie Greengard, David~W. Hogg,
  and Michael O'Neil.
\newblock Fast {{Direct Methods}} for {{Gaussian Processes}}.
\newblock {\em arXiv:1403.6015 [astro-ph, stat]}, March 2014.

\bibitem{golovin_google_2017}
Daniel Golovin, Benjamin Solnik, Subhodeep Moitra, Greg Kochanski, John Karro,
  and D.~Sculley.
\newblock Google {{Vizier}}: {{A Service}} for {{Black}}-{{Box Optimization}}.
\newblock In {\em Proceedings of the 23rd {{ACM SIGKDD International
  Conference}} on {{Knowledge Discovery}} and {{Data Mining}}}, KDD '17, pages
  1487--1495, New York, NY, USA, 2017. {ACM}.

\end{thebibliography}

\section*{Appendix: Tunable parameters of the BOP and FuBar VC algorithms}

\begin{tabular*}{1\columnwidth}{@{\extracolsep{\fill}}c>{\raggedright}p{0.6\columnwidth}}
\toprule 
Parameter & Description\tabularnewline
\midrule
\midrule 
$n_{\mathrm{cand}}$ & Number of local maxima to find in each Bayesian sampling step. \tabularnewline
\midrule 
$n_{\mathrm{poll}}$ & Number of random poll points candidates to find in each poll step \tabularnewline
\midrule 
$l_{\mathrm{poll}}$ & Size of poll step relative to estimated length scales\tabularnewline
\midrule 
$\rho$ & Minimal fraction of estimated noise level used for variance control
of the predicted standard deviation \tabularnewline
\midrule 
$SEM_{\min}$ & Minimal absolute level used for variance control of the predicted
standard deviation\tabularnewline
\midrule 
$z$ & Exponent of power law barrier function in the FuBar VC algorithm\tabularnewline
\midrule 
$x_{\mathrm{a.tol.}}$ & Absolute tolerance level for maximization of sample functions\tabularnewline
\midrule 
$T_{\mathrm{MCMC}}$ & Number of MCMC steps to take for each GP hyper-parameters sample\tabularnewline
\midrule 
$v_{\mathrm{noise}}$ & Scale of prior distribution of observation noise ($\sigma$)\tabularnewline
\midrule 
$A^{2}$ & Scale of prior distribution of GP amplitude ($\theta_{0}^{2}$)\tabularnewline
\midrule 
$\alpha_{\mathrm{length}}$ & Shape parameter of prior distribution of length scales ($\theta_{k},$
$k=1,2,\dots,D$)\tabularnewline
\midrule 
$\lambda_{\mathrm{length}}$ & Scale of prior distribution of length scales ($\theta_{k},$ $k=1,2,\dots,D$)\tabularnewline
\midrule 
exclude\_edge\_points & Should the algorithm exclude edge points (Boolean).\tabularnewline
\bottomrule
\end{tabular*}
\end{document}